\documentclass[aps,showpacs,showkeys,twocolumn]{revtex4}
\usepackage{mathrsfs}
\usepackage{amssymb}
\usepackage{amsmath}
\usepackage{bm}
\usepackage{graphics}
\usepackage{natbib}

\newcommand{\sss}[1]{{\scriptscriptstyle #1}}
\begin{document}
\title{Decoupling and antiresonance in electronic transport through a quantum dot chain
embodied in an Aharonov-Bohm interferometer}
\author{Yu Han$^{a,b}$}
\author{Weijiang Gong$^{a}$}\email[Corresponding author.
Fax: +086-024-8367-6883; phone: +086-024-8367-8327; Email address: ]
{weijianggong@gmail.com}
\author{Haina Wu$^{a}$}
\author{Guozhu Wei$^{a,c}$}

\affiliation{
a. College of Sciences, Northeastern University, Shenyang 110004, China \\
b. Department of Physics, Liaoning University, Shenyang 110036, China \\
c. International Center for Material Physics, Acadmia Sinica,
Shenyang 110015, China}
\date{\today}

\begin{abstract}
\textbf{Abstract} Electronic transport through a quantum dot chain
embodied in an Aharonov-Bohm interferometer is theoretically
investigated. In such a system, it is found that only for the
configurations with the same-numbered quantum dots side-coupled to
the quantum dots in the arms of the interferometer, some molecular
states of the quantum dot chain decouple from the leads. Namely, in
the absence of magnetic flux all odd molecular states decouple from
the leads, but all even molecular states decouple from the leads
when an appropriate magnetic flux is introduced. Interestingly, the
antiresonance position in the electron transport spectrum is
independent of the change of the decoupled molecular states. In
addition, when considering the many-body effect within the
second-order approximation, we show that the emergence of decoupling
gives rise to the apparent destruction of electron-hole symmetry. By
adjusting the magnetic flux through either subring, some molecular
states decouple from one lead but still couple to the other, and
then some new antiresonances occur.
\end{abstract}
 \maketitle

\bigskip

\section{Introduction}
During the past years, electronic transport through quantum-dot(QD)
systems has been extensively studied both experimentally and
theoretically. The atom-like characteristics of a QD, such as the
discrete electron levels and strong electron correlation, manifest
themselves by the experimental observations of Coulomb
blockade\cite{Thomas,Meirav,Sakaki,Zhang}, conductance
oscillation\cite{Kool}, and Kondo
effect\cite{Mahalu,Gores,Cronenwett,Heary} in electronic transport
through a QD. Therefore, a single QD is usually called an artificial
atom, and a mutually coupled multi-QD system can be regarded as an
artificial molecule. Thanks to the progress of nanotechnology, it
now becomes possible to fabricate a variety of coupled QD structures
with sizes to be smaller than the electron
coherence\cite{Xie,Shailos}. In comparison with a single QD, coupled
QD systems possess higher freedom in implementing some functions of
quantum devices, such as the QD cellular automata\cite{Loss1} and
solid-state quantum computation\cite{Loss2,Klein}.
\par
Motivated by an attempt to find some interesting electron transport
properties, recently many experimental and theoretical works have
become increasingly concerned about the electronic transport through
various multi-QD systems\cite{Tarucha,Yu,Vidan,Waugh,Amlani}.
According to the previous researches, the peaks of the linear
conductance spectra of coupled-QD systems reflect the eigenenergies
of the corresponding coupled QDs. On the contrary, the zero point of
the conductance, called antiresonance, originates from the
destructive quantum interference among electron waves passing
through different transmission paths. With respect to the coupled-QD
structures, the typical ones are the structures of the so-called
T-shaped QDs \cite{Iye,Sato,Ihn,Santos,Tor,Gong1}and the parallel
QDs \cite{Konign,Orellanan,Zhun,Sunn}( i.e., the QDs embodied in the
Aharonov-Bohm (AB) interferometer). A unique property of electron
transport through the T-shaped QD systems is that the antiresonance
points coincide with the eigenenergies of the side-coupled QDs,
which has also been observed experimentally\cite{Iye,Sato}. On the
other hand, the parallel-coupled QD systems, which offer two
channels for the electron tunneling, have also attracted much
attention. In such structures, with the adjustment of magnetic flux
the AB effect has been observed. Meanwhile, the appropriate
couplings between the molecular states of the coupled QDs and leads
can be efficiently adjusted, which gives rise to the tunable Fano
effect\cite{Orellanan,Zhun}. Moreover, under the condition of an
appropriate external field, some molecular states can decouple
completely from the leads, which is referred to as the formation of
bound states in continuum in some literature\cite{Orellana1}.
According to the previous researches, the existence of decoupling
plays a nontrivial role in the quantum interference of QD
structures, especially, it changes the property of the quantum
interference\cite{Bao}. Therefore, it is still desirable to clarify
the decoupling in electronic transport through some coupled-QD
structures.
\par
Since the development of nanotechnology, it is feasible to fabricate
the coupled QDs, in particular the QD chain, in the current
experiment\cite{Berry,Martin}. Thereby we are now theoretically
concerned with the electron transport properties of the this
structure, by considering it embodied in the AB interferometer. As a
result, we find that for the structures with the same-numbered QDs
side-coupled to the QDs in the two arms of the interferometer, some
molecular states of the QD chain decouple from the leads, and which
molecular states decouple from the leads is determined by the
adjustment of magnetic flux. Besides, in the case of the many-body
effect being considered, the existence of decoupling gives rise to
the destruction of electron-hole symmetry.

\section{model\label{theory}}
The coupled-QD structure we consider is illustrated in
Fig.\ref{structure}(a). The Hamiltonian that describes the
electronic motion in such a structure reads $H=H_{C}+H_{D}+H_{T}$,
in which $H_C$ is the Hamiltonian for the noninteracting electrons
in the two leads, $H_D$ describes the electron in the QD chain, and
$H_{T}$ denotes the electron tunneling between the leads and QDs.
They take the forms as follows.
\begin{eqnarray}
H_{C}&&=\underset{\sigma k\alpha\in L,R}{\sum }\varepsilon
_{k\alpha}c_{k\alpha\sigma}^\dag c_{k\alpha\sigma},\notag\\
H_{D}&&=\sum_{\sigma,m=1}^\sss{N}\varepsilon _{m}d_{m\sigma}^\dag
d_{m\sigma}+\sum_m U_m n_{m\uparrow}n_{m\downarrow}\notag\\
&&+\sum_{\sigma,m=1}^{N-1}(t_md^\dag_{m+1\sigma}d_{m\sigma}
+{\mathrm {H.c.}}),\notag\\
H_{T} &&=\underset{k\alpha\sigma}{\sum }( V_{\alpha
j}d_{j\sigma}^\dag c_{k\alpha\sigma}+V_{\alpha
j+1}d_{j+1\sigma}^\dag c_{k\alpha\sigma}+{\mathrm {H.c.}}),\notag\\
\end{eqnarray} where $c_{k\alpha\sigma}^\dag$ $(
c_{k\alpha\sigma})$ is an operator to create (annihilate) an
electron of the continuous state $|k,\sigma\rangle$ in lead-$\alpha$
with $\sigma$ being the spin index, and $\varepsilon _{k\alpha}$ is
the corresponding single-particle energy. $d^{\dag}_{m\sigma}$
$(d_{m\sigma})$ is the creation (annihilation) operator of electron
in QD-$m$, $\varepsilon_m$ denotes the electron level in the
corresponding QD, $t_m$ is the interdot hopping coefficient, and
$U_m$ represents the intradot Coulomb repulsion.
$n_{m\sigma}=d_{m\sigma}^\dag d_{m\sigma}$ is the electron number
operator in QD-$m$. We assume that only one level is relevant in
each QD and the value of $\varepsilon _{m}$ is independent of $m$,
i.e, $\varepsilon _{m}=\varepsilon_0$. In the expression of $H_{T}$,
the sequence numbers of the two QDs in the interferometer arms are
taken as $j$ and $j+1$, and $V_{\alpha j}$ and $V_{\alpha j+1}$ with
$\alpha=L, R$ denotes the QD-lead coupling coefficients. We adopt a
symmetric QD-lead coupling configuration which gives that
$V_{Lj}=Ve^{i\phi_L/2}$, $V_{Lj+1}=Ve^{-i\phi_L/2}$,
$V_{Rj}=Ve^{-i\phi_R/2}$, and $V_{Rj+1}=Ve^{i\phi_R/2}$ with $V$
being the QD-lead coupling strength. The phase shift $\phi_\alpha$
is associated with the magnetic flux $\Phi_\alpha$ threading the
system by a relation $\phi_\alpha=2\pi\Phi_\alpha/\Phi_{0}$, in
which $\Phi_{0}=h/e$ is the flux quantum.
\par
To study the electronic transport properties of such a structure,
the linear conductance at zero temperature is obtained by the
Landauer-B\"{u}ttiker formula
\begin{equation}
\mathcal {G}=\frac{e^{2}}{h}\sum_\sigma
T_\sigma(\omega)|_{\omega=\varepsilon_F}.\label{conductance}
\end{equation}
$T(\omega)$ is the transmission function, in terms of Green function
which takes the form as\cite{Meir1,Jauho}
\begin{equation}
T_\sigma(\omega)=\mathrm
{Tr}[\Gamma^LG^r_\sigma(\omega)\Gamma^RG^a_\sigma(\omega)],\label{transmission}
\end{equation}
where $\Gamma^L$ is a $2\times 2$ matrix, describing the strength of
the coupling between lead-L and the QDs in the interferometer arms.
It is defined as $[\Gamma^{L}]_{ll'}=2\pi
V_{\sss{L}l}V^*_{\sss{L}l'}\rho_\sss{L}(\omega)$ ( $l,l'=[j,j+1]$).
We will ignore the $\omega$-dependence of $\Gamma^{L}_{ll'}$ since
the electron density of states in lead-L, $\rho_\sss{L}(\omega)$,
can be usually viewed as a constant. By the same token, we can
define $[\Gamma^R]_{ll'}$. In fact, one can readily show that
$[\Gamma^L]_{ll}=[\Gamma^R]_{ll}$ in the case of identical QD-lead
coupling, hence we take $\Gamma=[\Gamma^L]_{ll}=[\Gamma^R]_{ll}$ to
denote the QD-lead coupling function. In Eq. (\ref{transmission})
the retarded and advanced Green functions in Fourier space are
involved. They are defined as follows:
$G_{ll',\sigma}^r(t)=-i\theta(t)\langle\{d_{l\sigma}(t),d_{l'\sigma}^\dag\}\rangle$
and
$G_{ll',\sigma}^a(t)=i\theta(-t)\langle\{d_{l\sigma}(t),d_{l'\sigma}^\dag\}\rangle$,
where $\theta(x)$ is the step function. The Fourier transforms of
the Green functions can be performed via
$G_{ll',\sigma}^{r(a)}(\omega)=\int^{\infty}_{-\infty}
G_{ll',\sigma}^{r(a)}(t)e^{i\omega t}dt$. These Green functions can
be solved by means of the equation-of-motion method\cite{Liu,Liu2}.
By a straightforward derivation, we obtain the retarded Green
functions which are written in a matrix form as
\begin{eqnarray}
G^r_\sigma(\omega)=\left[\begin{array}{cc}
g_{j\sigma}(z)^{-1} & -t_j+i\Gamma_{j,j+1}\\
-t^*_j+i\Gamma_{j+1,j}& g_{j+1\sigma}(z)^{-1}\\
\end{array}\right]^{-1}\ \label{green},
\end{eqnarray}
with $z=\omega+i0^+$ and $\Gamma_{ll'}={1 \over
2}([\Gamma^L]_{ll'}+[\Gamma^R]_{ll'})$.
$g_{l\sigma}(z)=[(z-\varepsilon_{l})S_{l\sigma}-\Sigma_{l\sigma}+i\Gamma_{ll}]^{-1}$,
is the zero-order Green function of the QD-$l$ unperturbed by
QD-$l'$, in which the selfenergies
\begin{eqnarray}&&\Sigma_{j\sigma}=\frac{t_{j-1}^2}{(z-\varepsilon_{j-1})S_{j-1\sigma}
-\frac{t^2_{j-2}}{(z-\varepsilon_{j-2})S_{j-2\sigma}-\ddots\frac{t_{2}^2}{(z-\varepsilon_{2})S_{2\sigma}
-\frac{t^2_{1}}{(z-\varepsilon_{1})S_{1\sigma}}}}}\notag\\
&&\Sigma_{j+1\sigma}=\frac{t_{j+1}^2}{(z-\varepsilon_{j+1})S_{j+1\sigma}
-\frac{t^2_{j+2}}{(z-\varepsilon_{j+2})S_{j+2\sigma}-\ddots\frac{t_{N-1}^2}{(z-\varepsilon_{N-1})S_{N-1\sigma}
-\frac{t^2_{N}}{(z-\varepsilon_{N})S_{N\sigma}}}}}\notag
\end{eqnarray}
account for the laterally coupling of the QDs to QD-$j$ and
QD-$j+1$, respectively\cite{Liu}. The quantity
$S_{m\sigma}=\frac{z-\varepsilon_{m}-U_m}{z-\varepsilon_{m}-U_m+U_m\langle
n_{m\bar{\sigma}}\rangle}$ ($m\in[1,N]$) is the contribution of the
intradot Coulomb interaction up to the second-order
approximation\cite{Gong1}. In addition, the advanced Green function
can be readily obtained via a relation
$G^a_\sigma(\omega)=[G^r_\sigma(\omega)]^\dag$.

\par
It is easy to understand that in the noninteracting case, the linear
conductance spectrum of the coupled QD structure reflects the
eigenenergy spectrum of the ``molecule" made up of the coupled QDs.
In other words, each resonant peak in the conductance spectrum
represents an eigenenergy of the total QD molecule, rather than the
levels of the individual QDs. Therefore, it is necessary to
transform the Hamiltonian into the molecular orbital representation
of the QD chain. We now introduce the electron
creation(annihilation) operators corresponding to the molecular
orbits, i.e., $f_{m\sigma}^\dag\; (f_{m\sigma})$. By the
diagonalization of the single-particle Hamiltonian of the QDs, we
find the relation between the molecular and atomic representations
(here each QD is regarded as an ``atom"). It is expressed as
$[\bm{f}_\sigma^\dag]=[\bm{\eta}][\bm{d}_\sigma^\dag]$. The $N\times
N$ transfer matrix $[\bm{\eta}]$ consists of the eigenvectors of the
QD Hamiltonian. In the molecular orbital representation, the
single-particle Hamiltonian takes the form:
$H=\underset{k\sigma\alpha\in L,R}{\sum }\varepsilon _{\alpha
k}c_{\alpha k\sigma }^\dag c_{\alpha k\sigma}+\sum_{m=1, \sigma}e
_{m}f_{m\sigma}^\dag f_{m\sigma}+\underset{\alpha k\sigma}{\sum }
v_{\alpha m}f_{m\sigma}^\dag c_{\alpha k\sigma}+{\mathrm {h.c.}}$,
in which $e_m$ is the eigenenergy of the coupled QDs. The coupling
between the molecular state $|m\sigma\rangle$ and the state
$|k,\sigma\rangle$ in lead-$\alpha$ can be expressed as
\begin{equation}
v_{\alpha m}=V_{\alpha j}[\bm\eta]^\dag_{jm}+V_{\alpha
j+1}[\bm\eta]^\dag_{j+1,m}.\label{gamma}
\end{equation}
In the case of symmetric QD-lead coupling, the above relation can be
rewritten as $v_{\alpha
m}=V([\bm\eta]^\dag_{jm}+[\bm\eta]^\dag_{j+1,m}e^{i\phi_\alpha})$.
Fig.\ref{structure}(b) shows the illustration of the QD structure in
the molecular orbital representation. We here define
$\gamma^\alpha_{mm}=2\pi v_{\alpha m} v^*_{\alpha
m}\rho_\alpha(\omega)$ which denotes the strength of the coupling
between the molecular state $|m\sigma\rangle$ and the leads.

\section{Numerical results and discussions \label{result2}}

With the theory in the above section, we can perform the numerical
calculation to investigate the linear conductance spectrum of this
varietal parallel double-QD structures, namely, to calculate the
conductance as a function of the incident electron energy. Prior to
the calculation, we need to introduce a parameter $t_0$ as the unit
of energy.
\par
We choose the parameter values $t_{m}=\Gamma=t_{0}$ for the QDs to
carry out the numerical calculation. And $\varepsilon_{0}$, the QD
level, can be shifted with respect to the Fermi level by the
adjustment of gate voltage experimentally. Typically, the case of
$\phi_L=\phi_R=\phi$ are first considered. Fig.\ref{QD2} shows the
linear conductance spectra ($\cal{G}$ versus $\varepsilon_0$) for
several structures with the QD number $N=2$ to $4$. It is obvious
that the 2-QD structure just corresponds to the parallel double QDs
with interdot coupling mentioned in some previous
works\cite{Orellanan,Zhun}. Its conductance spectrum presents a
Breit-Wigner lineshape in the absence of magnetic flux, as shown in
Fig.2(a). Such a result can be analyzed in the molecular orbital
representation. Here the $[\bm\eta]$ matrix, takes a form as
$[\bm\eta]={1 \over\sqrt{2}}\left[\begin{array}{cc}
-1& 1\\
1& 1\\
\end{array}\right]$, presenting the relation
between the molecular and `atomic' representations. Then with the
help of Eq. (\ref{gamma}) one can find that here the bonding state
completely decouples from the leads and only the antibonding state
couples to the leads, which leads to the appearance of the
Breit-Wigner lineshape in the conductance spectrum. On the other
hand, when introducing the magnetic flux with $\phi=\pi$, we can see
that the decoupled molecular state is changed as the antibonding
state, as exhibited by the dashed line in Fig.\ref{QD2}(a). In such
a case, only the bonding state couples to the leads and the
conductance profile also shows a Breit-Wigner lineshape.

\par
In Fig.\ref{QD2}(b) the conductance curves as a function of gate
voltage are shown for the 3-QD structure. Obviously, there exist
three conductance peaks in the conductance profiles and no decoupled
molecular state appears. We can clarify this result by calculating
$v_{\alpha
m}=V([\bm\eta]^\dag_{jm}+[\bm\eta]^\dag_{j+1,m}e^{i\phi})$. Via such
a relation, one can conclude that $v_{\alpha m}$ is impossible to be
equal to zero in this structure regardless of the adjustment of
magnetic flux. Thus one can not find the decoupled molecular states,
the state-lead coupling may be relatively weak, though. Just as
shown in Fig.\ref{QD2}(b), in the absence of magnetic flux the
distinct difference of the couplings between the molecular states
and leads offer the `more' and `less' resonant channels for the
quantum interference. Then the Fano effect occurs and the
conductance profile presents an asymmetric lineshape. In addition,
the Fano lineshape in the conductance spectrum is reversed by tuning
the magnetic flux to $\phi=\pi$, due to the modulation of magnetic
flux on $v_{\alpha m}$.

\par
When the QD number increases to $N=4$, there will be two
configurations corresponding to this structure, i.e, the cases of
$j=1$ and $j=2$. As a consequence, the conductance spectra of the
two structures remarkably differ from each other. With respect to
the configuration of $j=1$, as shown in Fig.\ref{QD2}(c), the
electron transport properties presented by the conductance spectra
are similar to those in the case of the 3-QD structure, and there is
also no existence of decoupled molecular states. However, for the
case of $j=2$, as shown in Fig.\ref{QD2}(d), it is clear that in the
absence of magnetic flux, there are two conductance peaks in the
conductance spectrum, which means that the decoupling phenomenon
comes into being. Alternatively, in the case of $\phi=\pi$, there
also exist two peaks in the conductance profile. But the conductance
peaks in the two cases of $\phi=0$ and $\pi$ do not coincide with
one another. We can therefore find that in this structure, when
$\phi=n\pi$ the decoupling phenomena will come about, and the
adjustment of magnetic flux can effectively change the appearance of
decoupled molecular states. By a further calculation and focusing on
the conductance spectra, we can understand that in the case of
$\phi=2n\pi$, the odd (first and third) molecular states of the
coupled QDs decouple from the leads; In contrast, the even (second
and fourth) molecular states of the QDs will decouple from the leads
if $\phi=(2n-1)\pi$. Additionally, in Fig.\ref{QD2}(d) it shows that
the conductance always encounters its zero when the level of the QDs
is the same as the Fermi level of the system, which is irrelevant to
the tuning of magnetic flux from $\phi=0$ to $\pi$.

\par
In order to obtain a clear physics picture about decoupling, we
analyze this problem in the molecular orbital representation. By
solving the $[\bm\eta]$ matrix and borrowing the relation $v_{\alpha
m}=V_{\alpha j}[\bm\eta]^\dag_{jm}+V_{\alpha
j+1}[\bm\eta]^\dag_{j+1,m}$, it is easy to find that in the case of
zero magnetic flux, $v_{\alpha 1}$ and $v_{\alpha 3}$ are always
equal to zero, which brings out the completely decoupling of the odd
molecular states from the leads. Unlike this case, when $\phi=\pi$
the values of $v_{\alpha 2}$ and $v_{\alpha 4}$ are fixed at zero,
and such a result leads to the even molecular states to decouple
from the leads. However, the underlying physics responsible for
antiresonance is desirable to clarify. We then analyze the electron
transmission by the representation transformation. We take the case
of $\phi=0$ as an example, where only two molecular states
$|2\sigma\rangle$ and $|4\sigma\rangle$ couple to the leads due to
decoupling. Accordingly, $|2\sigma\rangle$ and $|4\sigma\rangle$
might be called as well the bonding and antibonding states. As is
known, the molecular orbits of coupled double QD structures, e.g,
the well-known T-shaped QDs, are regarded as the bonding and
antibonding states. Therefore, by employing the representation
transformation
$[\bm{a}_\sigma^\dag]=[\bm{\beta}][\bm{f}_\sigma^\dag]$, such a
configuration can be changed into the T-shaped double-QD system (see
Fig.\ref{structure}(c)) of the Hamiltonian ${\cal
H}=\underset{k\sigma\alpha\in L,R}{\sum }\varepsilon _{\alpha
k}c_{\alpha k\sigma }^\dag c_{\alpha k\sigma}+\sum^2_{\sigma,
n=1}E_{n}a_{n\sigma}^\dag a_{n\sigma} +\tau_1 a_{2\sigma}^\dag
a_{1\sigma}+\underset{\alpha k\sigma}{\sum }
w_{\alpha1}a_{1\sigma}^\dag c_{\alpha k\sigma}+h.c.$. By a further
derivation, the relations between the structure parameters of the
two QD configurations can be obtained with $E_1=\varepsilon_0+t_0$,
$E_2=\varepsilon_0$, $\tau_1=t_0$, and $w_{\alpha1}=V_{\alpha1}$
respectively with
$[\bm\beta]={1\over\sqrt{2\sqrt{5}}}\left[\begin{array}{ccc}
-\sqrt{\sqrt{5}-1} & \sqrt{\sqrt{5}+1}\\
\sqrt{\sqrt{5}+1}& \sqrt{\sqrt{5}-1} \\
\end{array}\right].$
The 4-QD structure is then transformed into the T-shaped double QDs
with $\varepsilon_0$ being the level of dangling QD. Just as
discussed in the previous works\cite{Gong1}, in the T-shaped QDs
antiresonance always occurs when the dangling QD level is aligned
with the Fermi level of the system, one can then understand that in
this 4-QD system, the antiresonant point in the conductance spectrum
is consistent with $\varepsilon_0=\varepsilon_F=0$. When paying
attention to the $[\bm\eta]^\dag$ matrix, one will see that
$[\bm\eta]^\dag_{12}=[\bm\eta]^\dag_{43}$,
$[\bm\eta]^\dag_{22}=-[\bm\eta]^\dag_{33}$,
$[\bm\eta]^\dag_{32}=[\bm\eta]^\dag_{23}$, and
$[\bm\eta]^\dag_{42}=-[\bm\eta]^\dag_{13}$ for the 4-QD structure.
As a result, such relations give rise to $v_{\alpha
2}|_{\phi=0}=v_{\alpha 3}|_{\phi=\pi}$ and $v_{\alpha
4}|_{\phi=0}=v_{\alpha 1}|_{\phi=\pi}$. So, when $\phi=\pi$ the
magnetic flux reverses the lineshape of the conductance spectrum in
the case of $\phi=0$. Based on these properties, we can realize that
the quantum interference in the $\phi=\pi$ case is similar to that
in the case of $\phi=0$. Therefore, the antiresonant point in the
conductance spectra is independent of the adjustment of magnetic
flux.
\par
Similar to the analysis above, we can expect that in the structure
of with $t_m=t_0$ and $\varepsilon_m=\varepsilon_0$, when $N$ is
even and $j=\frac{N}{2}$ there must be the appearance of decoupled
molecular states. To be concrete, in the case of $\phi=2n\pi$, the
odd (first and third) molecular states of the coupled QDs decouple
from the leads, whereas the even (second and fourth) molecular
states of the QDs will decouple from the leads if $\phi=(2n-1)\pi$.
This expectation can be confirmed because of
$[\bm\eta]^\dag_{jm}=-[\bm\eta]^\dag_{j+1,m} (m\in \text{odd})$ and
$ [\bm\eta]^\dag_{jm}=[\bm\eta]^\dag_{j+1,m} (m\in \text{even})$ for
the QDs. The numerical results in Fig.\ref{6-8}, describing the
conductances of 6-QD and 8-QD structures, can support our
conclusion. Besides, with the help of the representation
transformation the antiresonance positions can be clarified by
transforming these structure into the T-shaped QD systems.
\par
Fig.\ref{manybody1} shows the calculated conductance spectra of the
double-QD structure by incorporating the many-body effect to the
second order and considering the uniform on-site energies of all the
QDs $U_m=U=2t_0$ as well as $4t_0$, respectively. It is seen that
the conductance spectra herein split into two groups due to the
Coulomb repulsion. But in each group the present decoupling
phenomenon and antiresonance are similar to those in the
noninteracting case. In addition, for each case ($U=2t_0$ or $4t_0$)
between the two separated groups a conductance zero emerges in the
conductance spectra. According to the previous discussions, when the
many-body terms are considered within the second-order
approximation, a pseudo antiresonance, resulting from the
electron-hole symmetry, should occur at the position of
$\varepsilon_0=-\frac{U}{2}$ where $\langle
n_{m\sigma}\rangle={1\over 2}$ \cite{Gong1,Liu,Liu2}. However, with
respect to such a situation, unlike the conventional electron-hole
symmetry, the position of such a conductance zero departs from
$\varepsilon_0=-\frac{U}{2}$ remarkably. Taking the structure of
double QDs as an example, in the absence of magnetic flux, this
conductance zero appears at the position of $\varepsilon_0=-{3\over
2}t_0$ when $U=2t_0$, but it presents itself at the point of
$\varepsilon_0=-3t_0$ in the case of $U=4t_0$, as shown in
Fig.\ref{manybody1}(a) and (c). Meanwhile, we can obtain
mathematically that around the point of
$\varepsilon_0=-\frac{U}{2}$, the average electron occupation number
$\langle n_{m\sigma}\rangle$ is not equal to ${1\over 2}$ any more.
Hence, the presence of decoupled molecular state destroys the
electron-hole symmetry. All these numerical results can be explained
as follows. For such a double-QD structure, one can understand that
when the many-body terms are considered within the second-order
approximation, the molecular levels are given by $\varepsilon_0-t_0,
\varepsilon_0+t_0, \varepsilon_0-t_0+U$, and $\varepsilon_0+t_0+U$,
respectively. So, when $\varepsilon_0=-{U\over 2}$, they distribute
symmetrically about the Fermi level of the system, which results in
$\langle n_{m\sigma}\rangle={1\over 2}$ and ${\cal G} =0$. But,
since the unique QD-lead coupling manner of our model, in the
absence of magnetic flux the bonding states ( with the levels
$\varepsilon_0-t_0$ and $\varepsilon_0-t_0+U$) completely decouple
from the leads and only the other two states ( $\varepsilon_0+t_0$
and $\varepsilon_0+t_0+U$ ) provide the channels for the electron
transport. Obviously, under the condition of $\varepsilon_0=-{U\over
2}$ the levels $\varepsilon_0+t_0$ and $\varepsilon_0+t_0+U$ do not
distribute symmetrically about the Fermi level of the system, thus
the electron-hole symmetry is broken, which shows itself as the
result of $\langle n_{m\sigma}\rangle\neq{1\over 2}$. Alternatively,
in the case of $\phi=\pi$, only the bonding states couple to the
leads, which also destroys the electron-hole symmetry, corresponding
to the results in Fig.\ref{manybody1}(b) and (d). The case of 4-QD
structure, as shown Fig.\ref{manybody2}, can also be clarified based
on such an approach.
\par
Next we turn to focus on the situation of $\phi_\alpha=n\pi$ and
$\phi_{\alpha'}\neq n\pi$. By virtue of Eq.(\ref{gamma}) we can
expect that for such a case, some molecular states of the QDs will
decouple from lead-$\alpha$ but they still couple to lead-$\alpha'$.
In Fig. \ref{phiLR} it shows the research on the electron transport
within the double-QD and four-QD $j=2$ structures. For the double-QD
structure, it is obvious that in the case of $\phi_L=0$ and
$\phi_R=0.5\pi$ both the bonding and antibonding states couple to
lead-R but the bonding state decouples from lead-L. The
corresponding numerical result is shown in Fig.\ref{phiLR}(a).
Clearly, due to the decoupling of the bonding state from lead-L,
antiresonance occurs at the position of $\varepsilon_0=t_0$. This
means that in this case antiresonance will come about when the level
of such a decoupled molecular state is consistent with the Fermi
level, corresponding to the discussions in the previous
works\cite{Liu2}. In the case of $\phi_L=0$ and $\phi_R=\pi$, the
bonding state decouples from lead-$L$ with the antibonding state
decoupled from lead-R. So herein there is no channel for the
electron tunneling and the conductance is always equal to zero,
despite the shift of QD level, corresponding to the dotted line in
Fig.\ref{phiLR}(a). On the other hand, by fixing $\phi_R=\pi$ and
increasing $\phi_L$ to $0.5\pi$, one will see that the decoupling of
antibonding state from lead-R results in the antiresonance at the
point of $\varepsilon_0=-t_0$. With regard to the 4-QD $j=2$
structure, the decoupling-induced antiresonance is also remarkable
in the case of $\phi_\alpha=n\pi$ and $\phi_{\alpha'}\neq n\pi$. As
shown in Fig.\ref{phiLR}(b) there are two kinds of antiresonant
points in such a case: One originates from the quantum interference
between the coupled molecular states, and the other is caused by the
decoupling of some molecular states. With respect to the many-body
effect, as shown in Fig. \ref{phiLR}(c), apart from the appearance
of two groups in the conductance spectra, the electron-hole symmetry
remains, which is due to that there is no completely decoupling of
the molecular states from the leads in such a case.

\par
In Fig.\ref{infinite} the linear conductances of the semi-infinite
and infinite QD chains are presented as a function of gate voltage.
As shown in Fig.\ref{infinite}, regardless of the semi-infinite or
infinite QD chains, no conductance peak is consistent with any
eigenlevel since the molecular states of the QDs become a continuum
in such a case. In the case of continuum, although some molecular
states decouple from the leads, it can not affect the electron
transport for reasons that the electron transmission paths can not
be differentiated. Thus no antiresonance appears in the conductance
spectra. However, when investigating the influence of the difference
between $\phi_L$ and $\phi_R$ on the electron transport, we find
that in the situation of $\phi_L=0$ and $\phi_R=0.5\pi$ the
conductance of the infinite QD chain encounters its zero at the down
side of energy band, as shown in Fig.\ref{infinite}(d). Such a
result can be explained as follows. The coupling of a semi-infinite
QDs to QD-$l$ indeed brings out the self-energy
$\Sigma_{l\sigma}=\frac{1}{2}(-\varepsilon_0-i\sqrt{4t_0^2-\varepsilon_0^2})$.
Then at both sides of the energy band (i.e.,
$\varepsilon_0=\pm2t_0$), the structure is just transformed into a
new double-QD configuration with
$\varepsilon_1=\varepsilon_2={\varepsilon_0\over 2}=t_0$. As a
result, in the case of $\varepsilon_0=2t_0$ the bonding state of
such a new double-QD structure, which has decoupled from lead-L
since $\phi_L=0$ and $\phi_R=0.5\pi$ ( as discussed in the previous
paragraph ), is aligned with the Fermi level, thereupon, the
electron tunneling here presents antiresonance. Alternatively, for
the case of $\phi_L=0.5\pi$ and $\phi_R=\pi$ a similar reason gives
rise to antiresonance at the position of $\varepsilon_0=-2t_0$. In
addition, it is apparent that in the case of $\phi_L=0$ and
$\phi_R=\pi$ the conductance is always fixed at zero. This is
because that in such a situation any molecular state coupled to
lead-$\alpha$ is inevitable to decouple from lead-$\alpha'$, though
the molecular states of the QDs is continuum. Thus, there is still
no channel for the electron transport.

\section{summary}
With the help of nonequilibrium Green function technique, the
electron transport through a QD chain embodied in an AB
interferometer has been theoretically investigated. It has been
found that for the configurations with the same-numbered QDs coupled
to the QDs in the interferometer arms, in the case of $\phi=2n\pi$
all odd molecular states of the QDs decouple from the leads, but all
even molecular states decouple from the leads when the magnetic flux
phase factor is equal to $(2n-1)\pi$. With the increase of magnetic
flux from $(2n-1)\pi$ to $2n\pi$, the antiresonance position in the
electron transport spectrum is independent of the change of the
decoupled molecular states. By representation transformation, these
results are analyzed in detail and the quantum interference in these
structures are therefore clarified. When the many-body effect is
considered up to the second-order approximation, we showed that the
emergence of decoupling gives rise to the apparent destruction of
electron-hole symmetry. Finally, the cases of different magnetic
fluxes through the two subrings were studied, it showed that via the
adjustment of the magnetic flux through either subring, some
molecular states would decouple from one lead but still couple to
the other, which cause the occurrence of new antiresonances.
\par
At last, we would like to point out that the theoretical model in
the present work can also be regarded as a double-QD AB
interferometer with some impurities side-coupled to the QDs in its
arms\cite{Liu2,EPL}, thus the calculated results can mimic the
influence of impurity states on the electronic transport behaviors
in such a structure. Therefore, we anticipate that the present work
may be helpful for the related experiments.

\clearpage

\bigskip
\begin{figure}
 \caption{ (a) Schematic of the
QD chain with two neighboring QDs coupled to both leads. Two
magnetic fluxes $\Phi_L$ and $\Phi_R$ thread the subrings in the
structure. (b) An illustration of the couplings between the
molecular states of the QDs and the leads. (c) presents a T-shaped
QD structure.\label{structure}}
\end{figure}

\begin{figure}
 \caption{The linear
conductance spectra of N-QD chains with $N=2$ to 4. The structure
parameters take the values as $\Gamma=t_{m}=t_{0}$, with $t_0$ being
the unit of energy. \label{QD2}}
\end{figure}

\begin{figure} \caption{(a) The
conductances of 6-QD system with $j=3$. In (b) The conductances of
8-QD structure are shown in the case of $j=4$. \label{6-8}}
\end{figure}

\begin{figure}
 \caption{The linear
conductance spectra of double-QD structure with the many-body terms
being considered. \label{manybody1}}
\end{figure}

\begin{figure}
 \caption{In the
presence of many-body terms, the linear conductance spectra of 4-QD
structure with $U=2t_0$ to $4t_0$. \label{manybody2}}
\end{figure}

\begin{figure} \caption{The calculated
conductance spectra of the double-QD and 4-QD structures in the
cases of $\phi_\alpha=n\pi$ and $\phi_{\alpha'}\neq n\pi$.
 \label{phiLR}}
\end{figure}

\begin{figure} \caption{The conductances
of the semi-infinite and infinite QD chains in the absence or
presence of magnetic flux. \label{infinite}}
\end{figure}

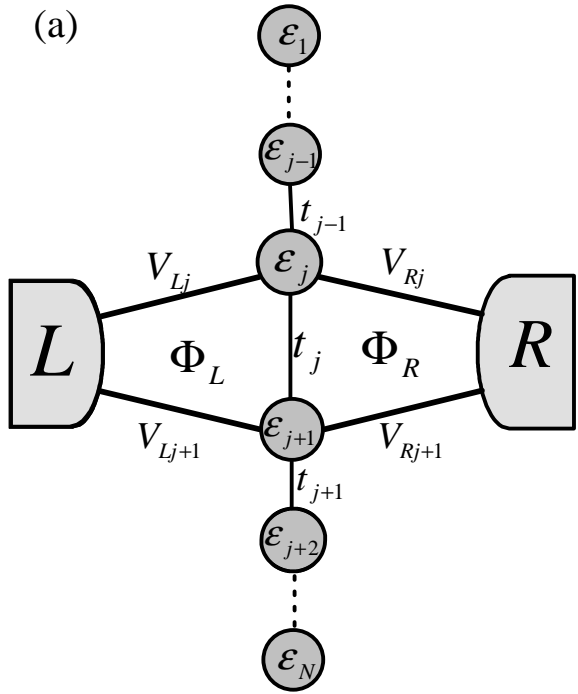
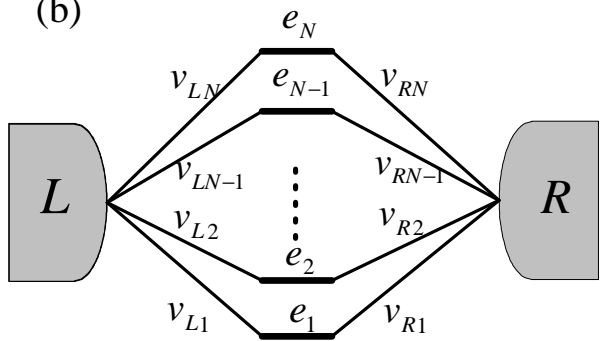
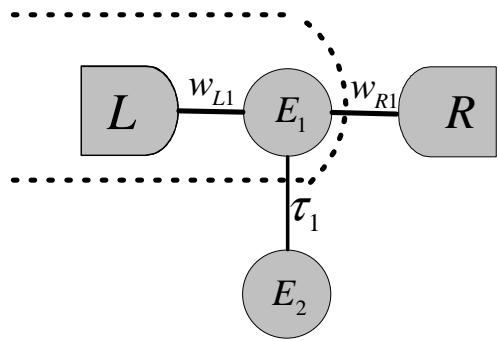
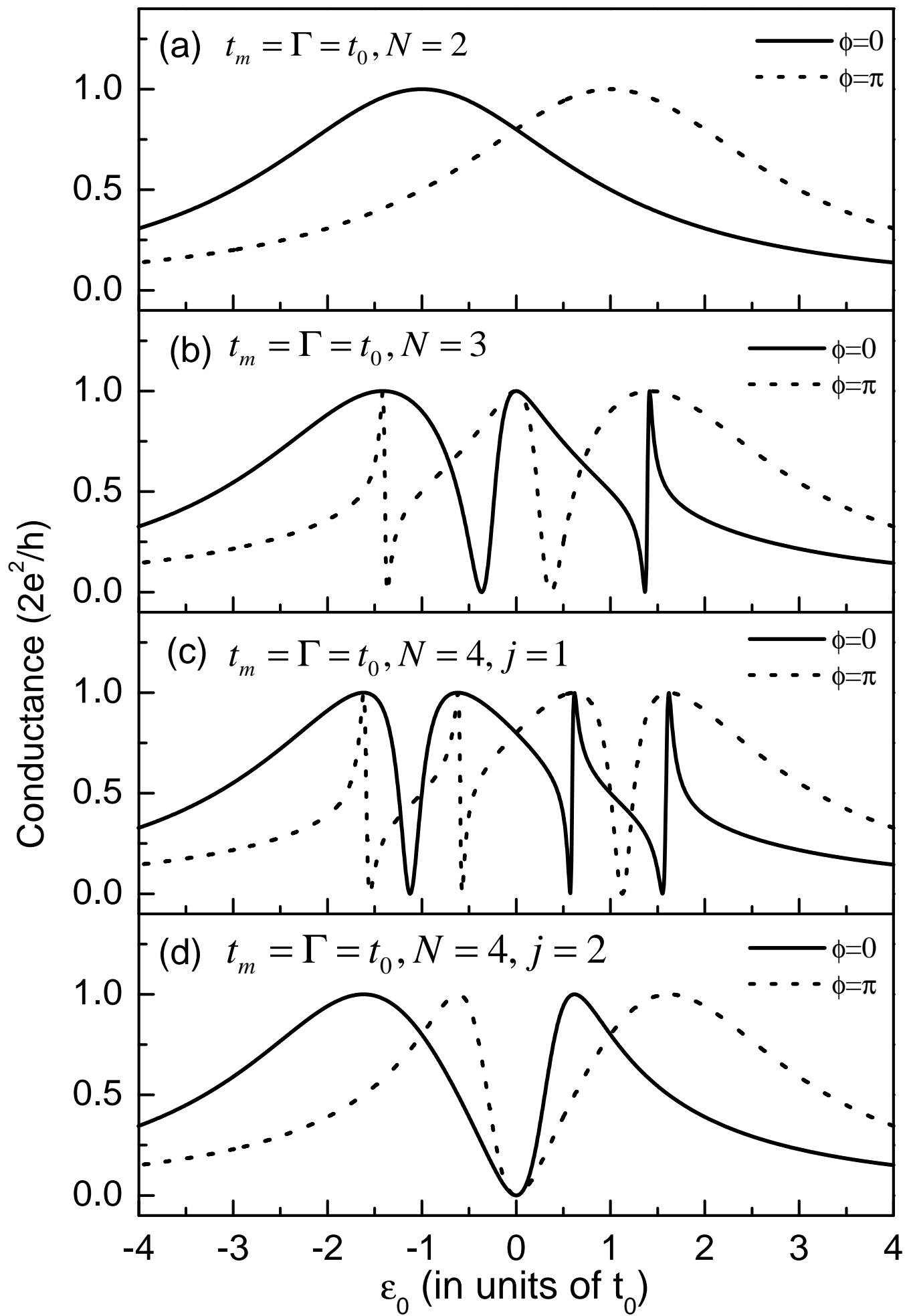
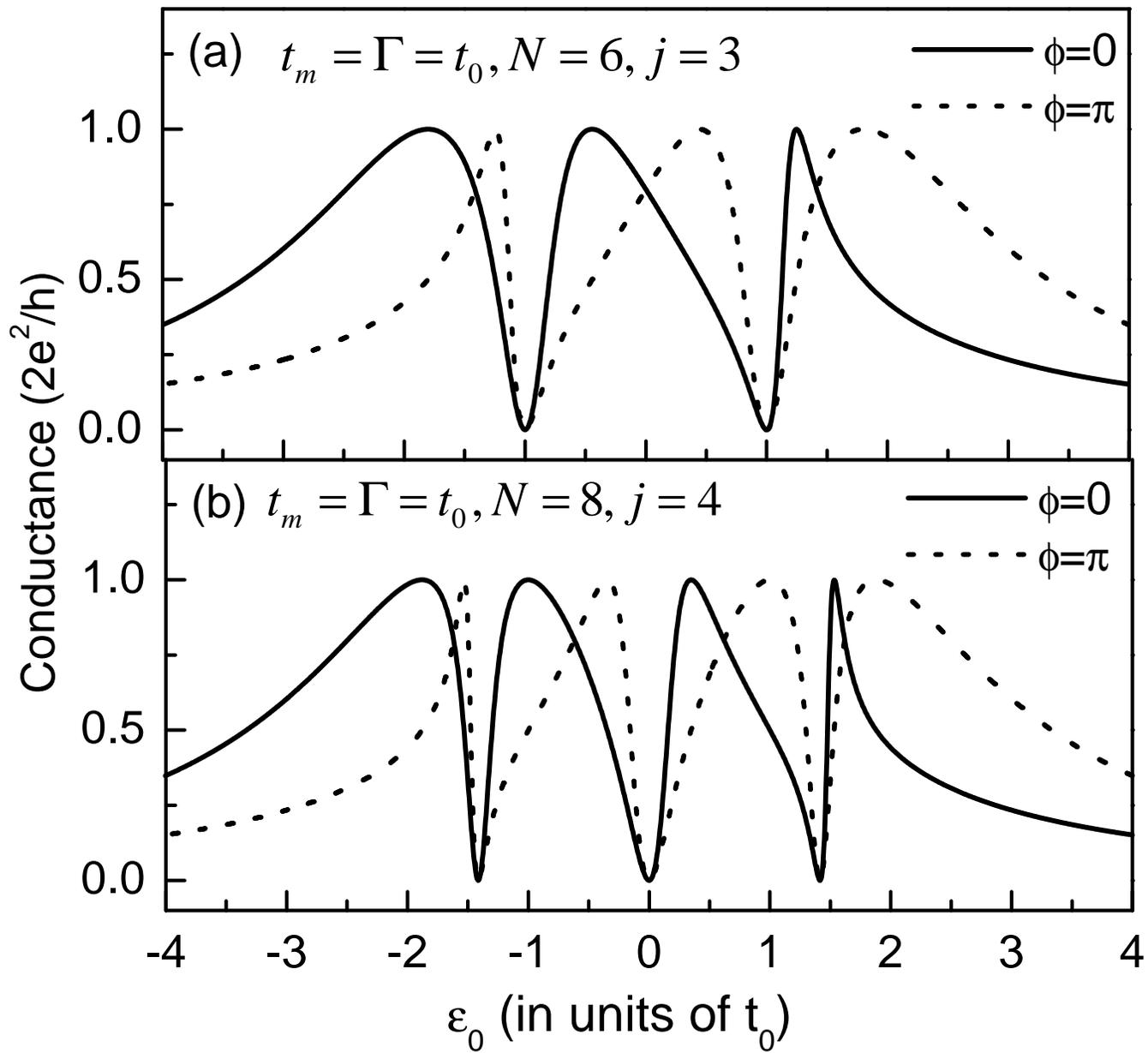
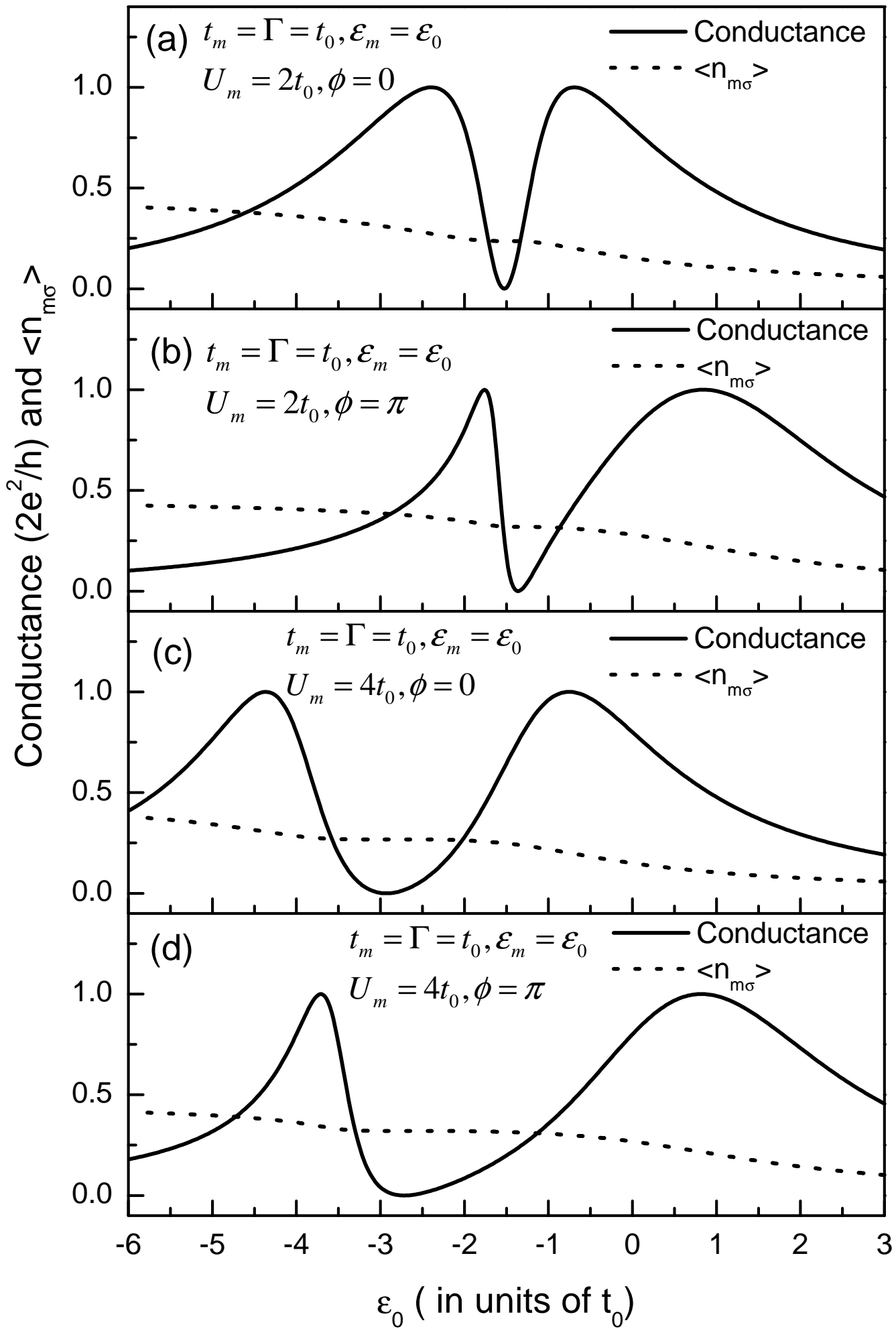
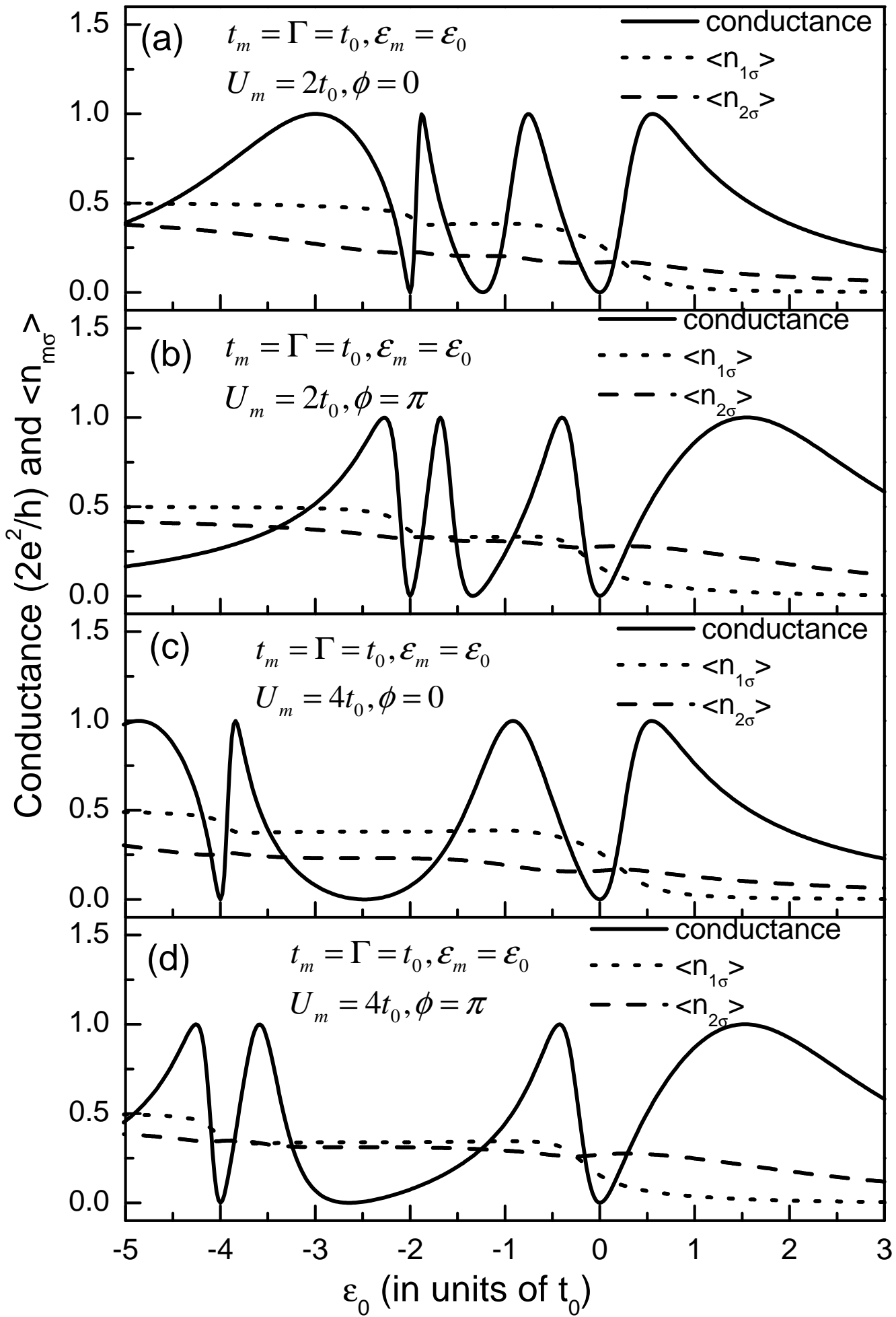
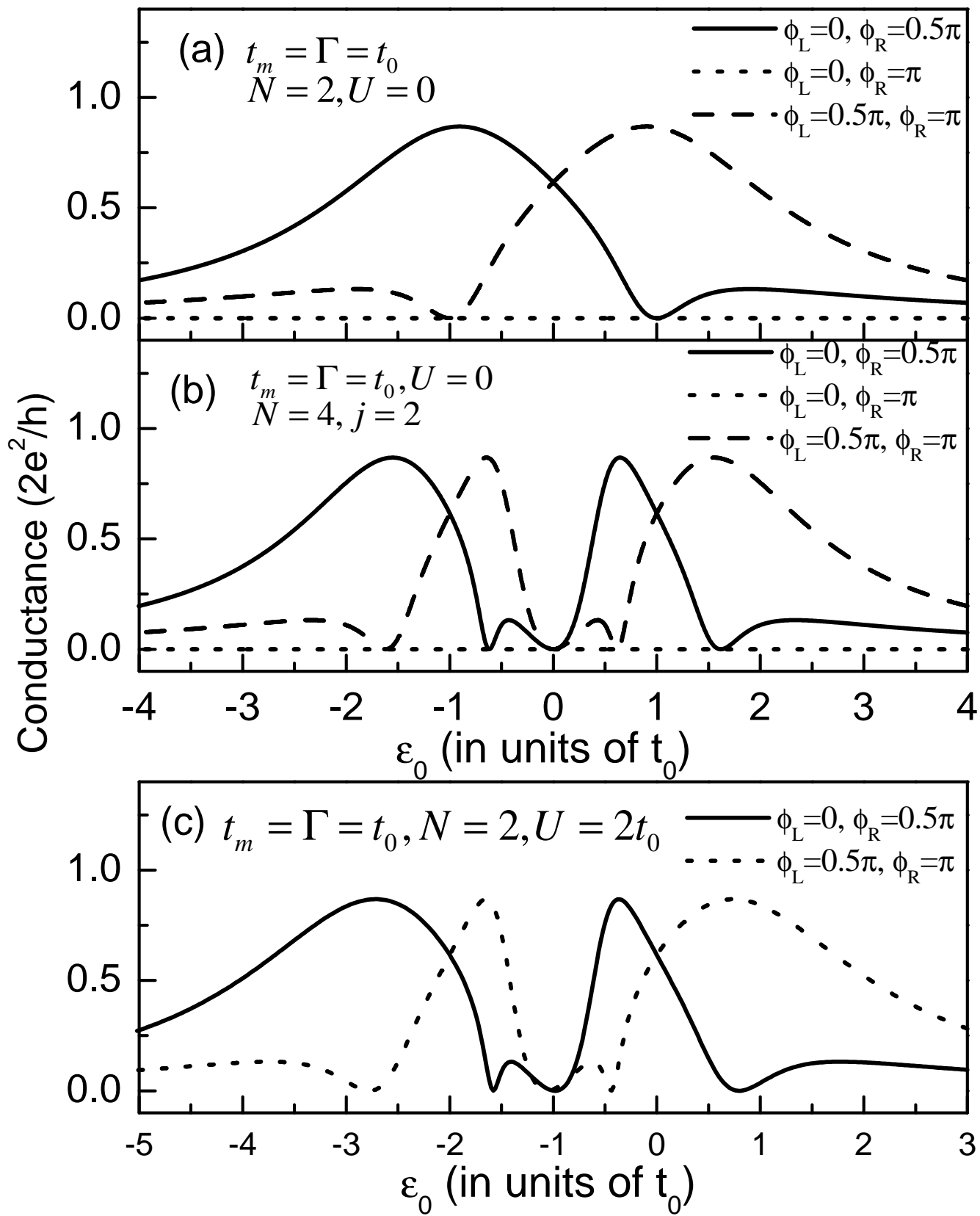
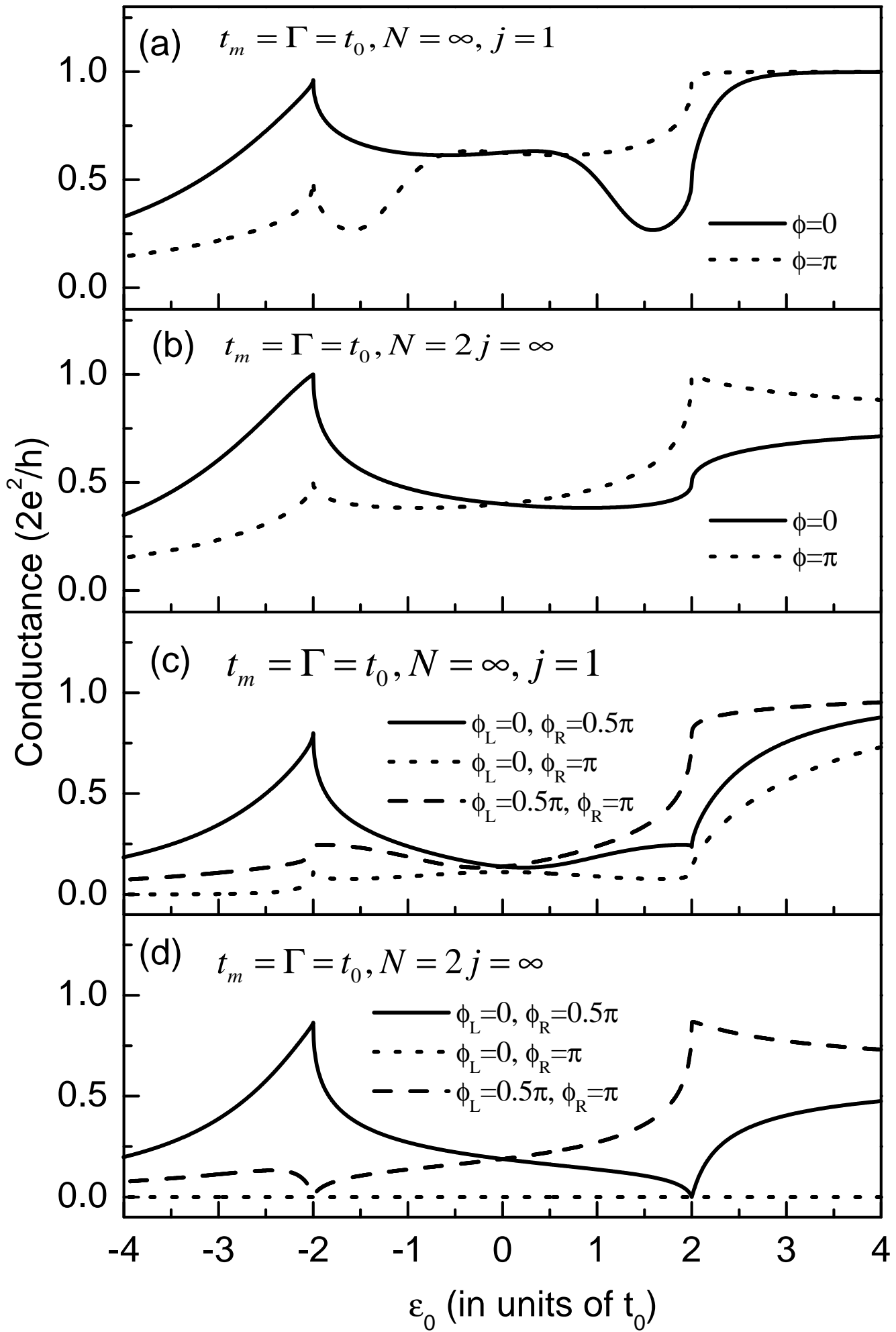
\end{document}